\documentstyle[epsfig,a4p,12pt,cite]{article}

\newcommand{\PNxxx}    
    
\def\eplemi{\mbox{$\mathrm{e^+ e^-}$}}

\def\Zz{\mbox{$\mathrm{Z}^0$}}

\def\Pip{\mbox{$\mathrm{ \pi^+}$}}

\def\Km{\mbox{$\mathrm{ K^-}$}}

\def\Dstar{\mbox{$\mathrm{ D^{\star+}}$}}
\def\Kstar{\mbox{$\mathrm{ K^{\star}}$}}

\def\Ups4s{\mbox{$\Upsilon(4S)$}}
\def\etal{\mbox{{\it et al.}}}

\def\GeVcc{\mbox{GeV/$c^2$}}
\def\GeVc{\mbox{GeV/$c$}}
\def\GeV{\mbox{GeV}}
\def\MeVc{\mbox{MeV/$c$}}
\def\MeVcc{\mbox{MeV/$c^2$}}

\newcommand{\inmath}[1] {\ifmmode#1\else$#1$\fi}
\newcommand{\definmath}[2] {\def#1{\ifmmode#2\else$#2$\fi}}
\definmath{\dEdx} {{\mathrm d}E/{\mathrm d}x}
\definmath{\PWpm} {\mathrm{W}^{\pm}}      
\definmath{\Pgtp} {\tau^{+}}        
\definmath{\Pgtm} {\tau^{-}}        
\definmath{\Pgtpm}   {\tau^{\pm}}         
\definmath{\Pgn}  {\nu}          
\definmath{\Pagn} {\overline{\nu}}     
\definmath{\Pq}      {\mathrm{q}}
\definmath{\Paq}  {\overline{\mathrm{q}}}
\definmath{\PQ}      {\mathrm{Q}}
\definmath{\PaQ}  {\overline{\mathrm{Q}}}
\definmath{\Pu}      {\mathrm{u}}
\definmath{\Pau}  {\overline{\mathrm{u}}}
\definmath{\Pd}      {\mathrm{d}}
\definmath{\Pad}  {\overline{\mathrm{d}}}
\definmath{\Ps}      {\mathrm{s}}
\definmath{\Pas}  {\overline{\mathrm{s}}}
\definmath{\Pc}      {\mathrm{c}}
\definmath{\Pac}  {\overline{\mathrm{c}}}
\definmath{\Pb}      {\mathrm{b}}
\definmath{\Pab}  {\overline{\mathrm{b}}}
\definmath{\Pt}      {\mathrm{t}}
\definmath{\Pat}  {\overline{\mathrm{t}}}
\definmath{\Pap}  {\overline{\mathrm{p}}}
\definmath{\Pan}  {\overline{\mathrm{n}}}
\definmath{\PaD}  {\overline{\mathrm{D}}}
\definmath{\PaDz} {\overline{\mathrm{D}}^{0}}
\definmath{\PaB}  {\overline{\mathrm{B}}}
\definmath{\PaBz} {\overline{\mathrm{B}}^{0}}
\definmath{\PsDpm}   {\mathrm{D}^{\pm}_{\mathrm{s}}}  
\definmath{\PcgLpm}  {\Lambda^{\pm}_{\mathrm{c}}}  
\definmath{\PDst} {\mathrm{D}^{*}}     
\definmath{\PKs} {\mathrm{K}^{0}_{\mathrm s}}     
\definmath{\PgLz} {\Lambda^{0}}        
\newcommand{\qqbar}  {\Pq\Paq}

\newcommand{\bbbar}  {\Pb\Pab}

\newcommand{\Ztobb}     {\Zz\to\bbbar}

\newcommand{\Gbb}    {\Gamma_{\mathrm b \overline{\mathrm b}}}
\newcommand{\Ghad}      {\Gamma_{\mathrm{had}}}
\newcommand{\GbbGhad}      {\Gbb/\Ghad}

\definmath{\GeV}  {\mathrm{GeV}}
\definmath{\GeVc} {{\mathrm{GeV}}\!/c}
\definmath{\GeVcc}   {{\mathrm{GeV}}\!/c^2}
\definmath{\MeV}  {\mathrm{MeV}}
\definmath{\MeVc} {{\mathrm{MeV}}\!/c}
\definmath{\MeVcc}   {{\mathrm{MeV}}\!/c^2}
\definmath{\MVm}  {\mathrm{MV}\!/\mathrm{m}}
\definmath{\keV}  {\mathrm{keV}}
\definmath{\keVcm}   {\mathrm{keV}\!/\mathrm{cm}}
\definmath{\kV}      {\mathrm{kV}}
\definmath{\km}      {\mathrm{km}}
\definmath{\meter}   {\mathrm{m}}
\definmath{\cm}      {\mathrm{cm}}
\definmath{\mm}      {\mathrm{mm}}
\definmath{\micron}  {\mu\mathrm{m}}
\definmath{\nm}      {\mathrm{nm}}
\definmath{\kg}      {\mathrm{kg}}
\definmath{\gram} {\mathrm{g}}
\definmath{\second}  {\mathrm{s}}
\definmath{\microsec}   {\mu\mathrm{s}}
\definmath{\degree}  {^\circ}
\definmath{\degC} {^\circ\mathrm{C}}
\definmath{\ohm}  {\Omega}
\definmath{\Mohm} {\mathrm{M}\Omega}
\definmath{\rad}  {\mathrm{rad}}
\definmath{\mrad} {\mathrm{mrad}}
\definmath{\nb}      {\mathrm{nb}}
\newcommand{\eqref}[1]  {(\ref{#1})}
\newcommand{\PhysLett}  {Phys.~Lett.}

\newcommand{\NIM} {Nucl.~Instr.\ Meth.}
\newcommand{\ZPhys}  {Z.~Phys.}

\newcommand{\OPALColl}  {OPAL Collab.}
\newcommand{\JADEColl}  {JADE Collab.}

\def\cent{\centerline}

\def\to{$\rightarrow$}

\def\plm{$\pm$}

\begin{document}
\begin{titlepage}
\begin{center}{\large   EUROPEAN ORGANIZATION FOR NUCLEAR RESEARCH
}\end{center}\bigskip
\begin{flushright}
       CERN-EP/99-179   \\ 17 December 1999
\end{flushright}

\vspace{2cm}
 
\begin{center}
{\Huge\bf Search for New Physics in Rare B Decays}
\end{center}
\vspace{1cm}
\begin{center}{\LARGE The OPAL Collaboration
}\end{center}
\vspace{1cm}
\vspace{.5cm}
\begin{center} 
\end{center}

\cent{\large\bf Abstract}
A search for the decay  B$^\pm$\to K$^\pm$K$^\pm\pi^\mp$ was
performed using data collected by the OPAL detector at LEP. 
These decays are strongly suppressed in the Standard Model but could
occur with a higher branching ratio in supersymmetric models, especially
in those with R-parity violating couplings.
No evidence for a signal was observed and a 90\% confidence level upper
limit of $1.29\times 10^{-4}$ was set for the branching ratio.

\vspace{2cm}

\cent {\large Submitted to Physics Letters B}
\vspace{1cm}

\end{titlepage} 
\begin{center}{\Large        The OPAL Collaboration
}\end{center}\bigskip
\begin{center}{
G.\thinspace Abbiendi$^{  2}$,
K.\thinspace Ackerstaff$^{  8}$,
P.F.\thinspace Akesson$^{  3}$,
G.\thinspace Alexander$^{ 22}$,
J.\thinspace Allison$^{ 16}$,
K.J.\thinspace Anderson$^{  9}$,
S.\thinspace Arcelli$^{ 17}$,
S.\thinspace Asai$^{ 23}$,
S.F.\thinspace Ashby$^{  1}$,
D.\thinspace Axen$^{ 27}$
% ...
}\end{center}\bigskip
\bigskip
$^{  1}$School of Physics and Astronomy, University of Birmingham,
Birmingham B15 2TT, UK
\newline
$^{  2}$Dipartimento di Fisica dell' Universit\`a di Bologna and INFN,
I-40126 Bologna, Italy
\newline
$^{  3}$Physikalisches Institut, Universit\"at Bonn,
D-53115 Bonn, Germany
\newline
% ...
$^{  8}$CERN, European Organisation for Particle Physics,
CH-1211 Geneva 23, Switzerland
\newline
$^{  9}$Enrico Fermi Institute and Department of Physics,
University of Chicago, Chicago IL 60637, USA
\newline
% ...
$^{ 16}$Department of Physics, Schuster Laboratory, The University,
Manchester M13 9PL, UK
\newline
$^{ 17}$Department of Physics, University of Maryland,
College Park, MD 20742, USA
\newline
% ...
$^{ 22}$Department of Physics and Astronomy, Tel Aviv University,
Tel Aviv 69978, Israel
\newline
$^{ 23}$International Centre for Elementary Particle Physics and
Department of Physics, University of Tokyo, Tokyo 113-0033, and
Kobe University, Kobe 657-8501, Japan
\newline
% ...
$^{ 27}$University of British Columbia, Department of Physics,
Vancouver BC V6T 1Z1, Canada
\newline
% ...
\eject
\section{Introduction}
\label{Sec:Intro}
Rare b decays offer an opportunity to discover new physics
beyond the Standard Model (SM). Many studies have been done in
recent years to predict flavor
changing neutral current (FCNC) processes, both within the SM and
beyond \cite{PSI}. One of these FCNC induced processes, b\to s$\gamma$,
has been measured \cite{PSII} and the
branching ratio found to be consistent with the SM prediction \cite{PSIII}. 
However, significant uncertainties still remain in both the
theoretically 
predicted 
branching ratio and the measurement. Thus it is hard to
conclude if this process shows signs of new physics \cite{JH}. This is also
true in most of the channels such as b\to s${\mathrm q\overline q}$\cite{PSIV} and
b\to s$\ell\overline \ell$\cite{PSV}, due to theoretical uncertainties.

\begin{figure}[h]
   \begin{center}\hspace*{-2.cm}\mbox{
        \epsfig{file=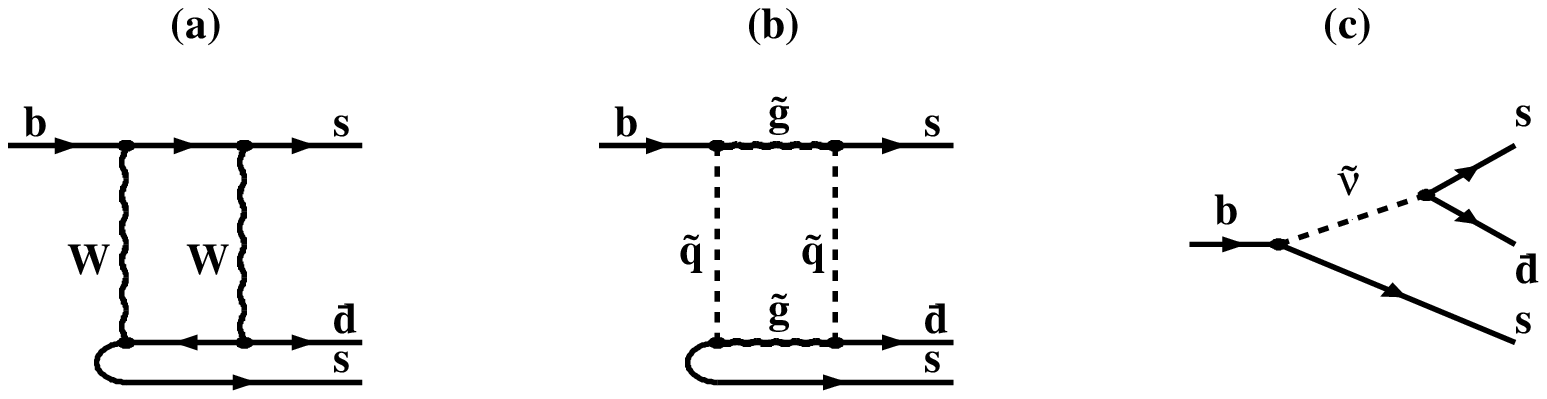}}
   \caption{b\to ss$\overline{\mathrm d}$ transition (a) SM, (b) MSSM, (c) MSSM with
     R-parity violating coupling.}
   \label{fig:diag}
\end{center}
\end{figure}

The process b\to ss${\mathrm\overline d}$, induced by a box diagram,
is predicted to be exceedingly
small in the SM (Figure \ref{fig:diag}(a)), of the order of $10^{-11}$
\cite{PS}. However, 
in the minimal supersymmetric standard model
(MSSM) \cite{MSSM}, this transition can be induced by the squark-gaugino (or
higgsino) box diagrams (Figure \ref{fig:diag}(b)) at a level of 
$10^{-7}-10^{-8}$. An alternative
mechanism for this channel in
supersymmetric models is through R-parity violating couplings
\cite{rparity} (Figure \ref{fig:diag}(c)).
These two possibilities appear to be the only ones that will produce 
significant 
enhancement of this decay within supersymmetric models\cite{PS}.
Two higgs doublet models could also induce this decay at branching
ratios significantly larger than in the Standard Model, for a
certain range of the parameters involved \cite{newPS}.

Typical exclusive processes of b\to ss${\mathrm \overline d}$ include 
B$^\pm$\to K$^\pm$ K$^0,$ which are difficult to separate from the standard
penguin process b\to ds${\mathrm \overline s}$.
Although the interference of these two sources of the final
state is crucial in the study of phenomena such as CP violation,
this channel is not suitable for a direct search for new
physics. However the decay 
B$^-$\to \Km\Km\Pip \footnote{charge conjugation is assumed
  throughout this paper}, either as a direct three-body decay or through a
\Kstar-like resonance, is a clear signature of this process.
This document describes the first search for the decay 
B$^-$\to \Km\Km\Pip. 
\vspace*{-2.cm}

\section{Hadronic Event Selection and Simulation}
 \label{Sec:EvtSel}
We used data collected at LEP by the OPAL detector \cite{OPAL} between
1990 and 1995 running at center-of-mass energies in the vicinity of
the \Zz\ peak.
Hadronic \Zz\ decays were selected using the number of charged tracks
and the visible energy in each event as in Reference \cite{evsel}. This
selection yielded 4.41 million hadronic events. 

Monte Carlo events were used to determine the selection efficiency,
for training of an artificial neural network (ANN) used in
selecting the final event sample (Section \ref{Sec:Analysis}), 
and for the determination of some of the systematic uncertainties
(Section \ref{Sec:Syst}).
To determine the selection efficiency we generated 100\thinspace 000  Monte
Carlo 
events of the process \Ztobb\ where one of the b quarks hadronised into a B$^-$
meson which subsequently decayed to \Km\Km\Pip. One sample 
was generated according to three-body decay phase space, while another
sample was generated
with angular distribution as expected from a weak decay matrix element. 
In addition,
we generated samples in which the B meson decayed via an intermediate \Kstar\
resonance. \Kstar(892), \Kstar(1680) and
\Kstar(2045) resonances were chosen as they all decay into \Km\Pip\ and cover the
entire spectrum of \Kstar\ resonances.

For optimisation of the selection of events and for some of the
studies of systematic uncertainties, we generated 4 million 5-flavour
hadronic \Zz\ decays (referred to as \qqbar\ Monte Carlo). 
All these samples were 
generated with the JETSET
7.4 Monte Carlo program \cite{jetset} with parameters tuned to the
OPAL data \cite{OPALtune}. The heavy quark
fragmentation was parameterised by the fragmentation
function of Peterson \etal\ \cite{peterson}, and all samples were processed with the 
OPAL detector simulation package~\cite{gopal}.

\section{Analysis Procedure}
\label{Sec:Analysis}

In each event, charged tracks and electromagnetic clusters not
associated to a charged track were combined into jets, using the
JADE algorithm with the E0 recombination scheme\cite{jade}.
Within this algorithm jets are defined by 
$y_{\mathrm {cut}}=0.04,$ where $y_{\mathrm {cut}}$ is defined in
Reference \cite{jade}.

The primary vertex of the event was reconstructed using the 
charged tracks in the event and the knowledge of the position and 
spread of the \eplemi\ collision point.

We searched the hadronic event sample for the decay B$^-$\to
\Km\Km\Pip\ by combining three charged tracks to form a B meson
candidate. All three track combinations were considered.
All tracks were required to have a momentum of at least
2 \GeVc\ and to be in the same jet. 
Two of the tracks were required to have the same charge and
were assigned the mass of a kaon. A third track, with an opposite charge, was
assumed to be the pion.  Tracks were required to satisfy selection
criteria based on the
measured rate of energy loss due to ionisation (\dEdx) \cite{dedx} 
as listed in Table 1.
These \dEdx\ selection criteria are 44\% efficient while rejecting
98.5\% of the background.

The three tracks were fitted to a common vertex and the decay
length, the distance from the \eplemi\ interaction point to the
reconstructed secondary vertex, was calculated. 
Candidates where the secondary vertex is in the hemisphere opposite to the
candidate's jet were rejected. This criterion left 55\% of the remaining
background events, but kept 96\% of the signal events. 

Since the hadronic data sample consisted mostly of
non-\bbbar\ events, we suppressed these events by means of a b-tagging
algorithm, based on reconstructed displaced secondary vertices. 
An artificial neural network with inputs based on
decay length significance, vertex multiplicity and invariant mass 
information \cite{VNN} was used to select
vertices with a high probability of coming from b hadron
decays. Events were accepted if any of the jets were tagged by the
neural network. The b-tagging selection was found to
be 79\% efficient, while rejecting 80\% of the remaining background.

\begin{table}
\begin{center}
\begin{tabular}{|l|c|c|} \hline
Selection &$\pi$ selection & K selection \\
\hline
$\dEdx_\pi$ & - & $<-1.29\sigma$ \\
$\dEdx_K$ &  $>1.15\sigma$    &$\pm1.55\sigma$\\
$\dEdx_p$ &  -  &  $>0$ \\
$N_{\dEdx}$  & $>20$ & $>20$\\
\hline
\end{tabular}
\caption{Summary of $\dEdx$ selection criteria: $\dEdx_{\pi(K)}$ is the
  difference between the measured value of the ionisation energy loss
  in the jet chamber and the expected value for $\pi$ (K) and
  $\sigma$ represents the expected standard deviation of the
  distribution. $N_{\dEdx}$ is the number of jet chamber hits
  used for $\dEdx$. The $\dEdx$
  cuts were chosen based on probability values.}
\label{tab:dedx}
\end{center}
\end{table}

\begin{figure}[]
   \begin{center}\hspace*{-.4cm}\mbox{
        \epsfig{file=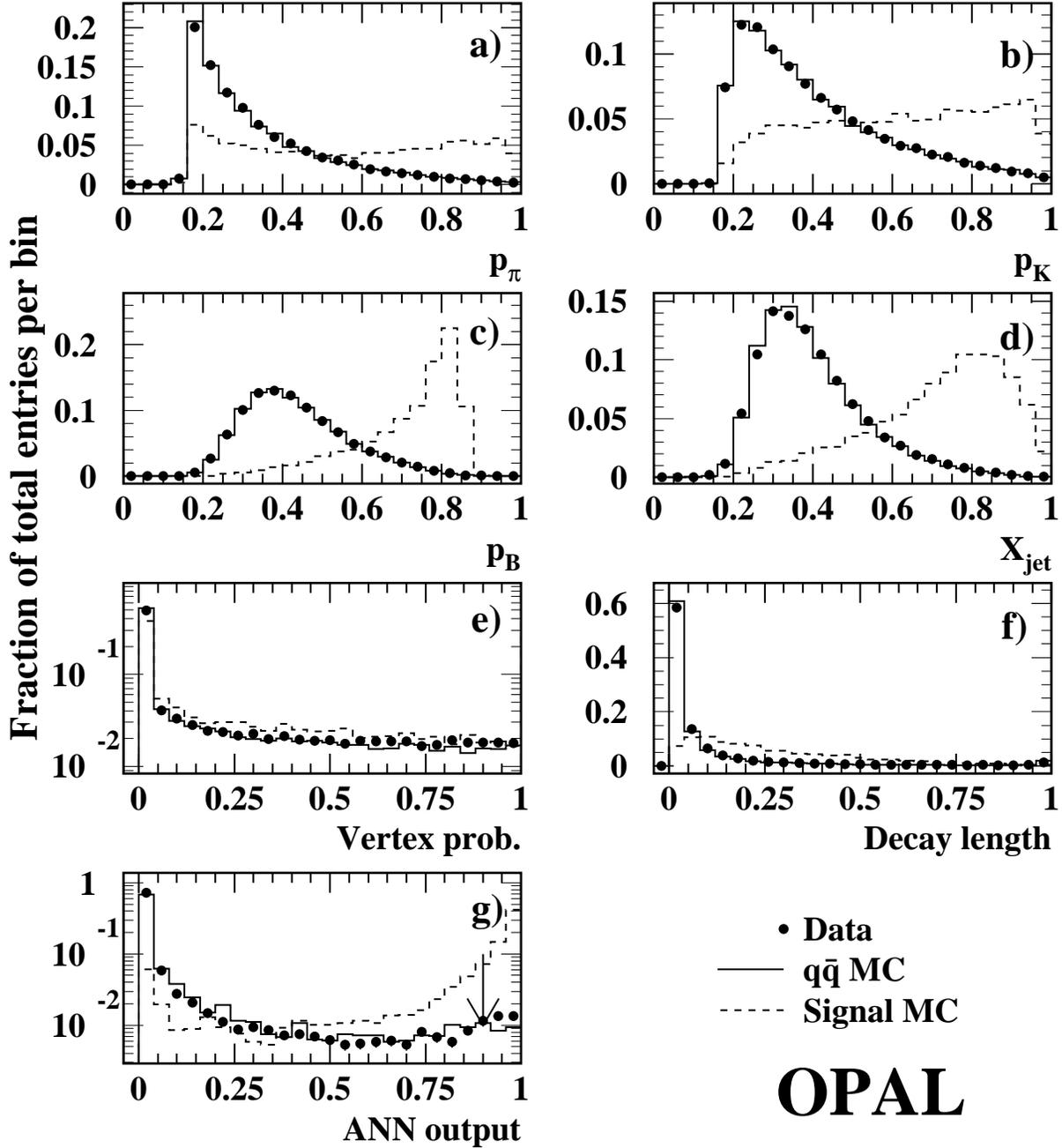}}
   \caption{Input variables to the artificial neural network (a-f) and
     output (g). The solid line represents the \qqbar\ Monte Carlo
     while the 
     dots represent the data. The dashed line shows
     the distribution of B$^-$\to K$^{\star0}$(892)\Km\ Monte
     Carlo events. The arrow in (g) shows the cut value.  All
     variables are normalised and are plotted after appropriate transformation
     to the range $[0-1]$.}
   \label{fig:nn}
\end{center}
\end{figure}

The final selection was based on an artificial neural network designed
to select B$^-$\to\Km\Km\Pip\ events while rejecting background
events. We used the JETNET 3.4 program \cite{jetnet} with a
feed-forward type net, trained with the back-propagation algorithm. 
The neural network used seven input parameters: the
momenta of the three tracks ($p_\pi$, $p_{\mathrm K}$); the B
candidate momentum ($p_{\mathrm B}$); the ratio of
B candidate energy to the jet energy ($X_{\mathrm jet}$); the decay
length; and the vertex probability, the probability of the three
tracks to originate from a common vertex which is calculated using the
track parameters. 
The neural network retains 74\% of
the signal events and rejects 97\% of background events when selecting
candidates with an ANN output above 0.9. Figure
\ref{fig:nn} shows the distributions of the artificial neural network
input parameters and the ANN output for candidates passing the above
criteria. While the data and the \qqbar\ Monte
Carlo show good agreement, one can clearly observe the
differences between these distributions and those of the signal Monte Carlo.

\begin{figure}[]
   \begin{center}\hspace*{-.4cm}\mbox{
        \epsfig{file=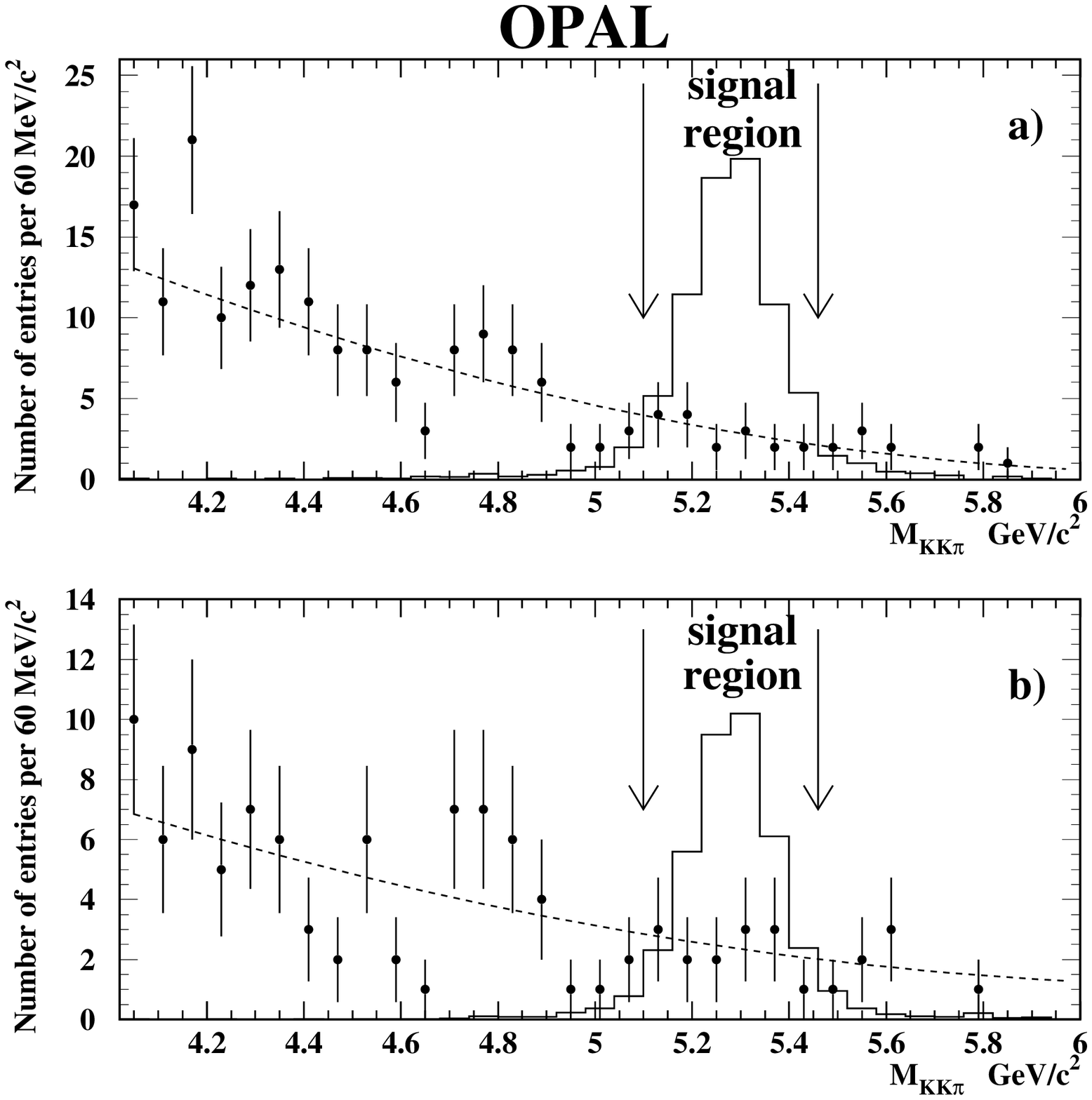}}
   \caption{Invariant mass distribution of the \Km\Km\Pip\ candidates
     after all selection criteria were applied a) via intermediate
     \Kstar\ resonance and b) with direct production. The 
     dots represent the data, the solid line shows the expected
     signal shape from Monte Carlo events after all the selection
     criteria were applied with arbitrary normalisation, and the
     dashed line is the expected background.}
   \label{fig:bmass}
\end{center}
\end{figure}

Candidates were accepted if their invariant mass was in the region:  
$5.10~\GeVcc~<$ $~M_{\mathrm KK\pi}~$ $<~5.46~\GeVcc,$ which
corresponds to twice the mass resolution around the nominal B$^-$ mass.
Only one candidate per jet was accepted, based on the largest neural
network output for candidates in a jet.
Figure \ref{fig:bmass} (a) shows the invariant mass distribution of the
\Km\Km\Pip\ candidates. No enhancement is seen in the signal region,
where 17 events were observed. 
Monte Carlo studies indicated that 88\% of the background
at this stage consists of \bbbar\ events.

If the decay chain B$^-$\to\Km\Km\Pip\ is assumed to be direct
(i.e., without an intermediate \Kstar\ resonance), then the mass of the
\Km\Pip\ system can be exploited to further reduce background where
the pion and one of the kaons are from the decay of a \Kstar\ resonance.
The mass of the \Km\Pip\ system was added as an input to the neural network, 
and the training procedure of the ANN was repeated.
Figure \ref{fig:bmass} (b) shows the invariant mass distribution of the
\Km\Km\Pip\ candidates passing the selection. Here too, no
enhancement is seen in the signal region and the observed 14 events
are used to determine an upper limit on the
branching ratio.

\subsection{Background Estimation}

The background to the process B$^-$\to\Km\Km\Pip\ was estimated
by fitting a second-order polynomial to the invariant mass of a
combinatorial background, obtained by releasing the ANN cut, and then 
normalising the shape to the 
mass side-bands of Figure \ref{fig:bmass} (4--5~\GeVcc\
and  5.6--6~\GeVcc). Monte Carlo studies indicated that the
background shape is not altered by this procedure. Alternatively, we
repeated this procedure by releasing each of the selection criteria
separately and by obtaining the shape from Monte Carlo. All the
alternative fits gave a consistent result. We also took the number of
events within the signal region in each of the above cases and scaled
it to the appropriate sample size. Here too, all estimates were
consistent. 

As we are setting upper limits, the conservative approach is to
estimate the number of signal events using the lowest background
estimate. While the fitted background shown in Figure \ref{fig:bmass} (a)
gave 18.8 events in the signal region, the lowest estimate 
was 17.5 events. The respective numbers for the direct production case
were 14.8 and 14.1 events.

\subsection{Limit Determination}

The above numbers were used to determine $N^{90}$, the 90\% C.L. upper limit
on the number of signal events. Using the formalism of reference
\cite{N90}, we obtained $N^{90}=7.8$ and $N^{90}=7.4$ events for the
resonant/direct decay, respectively.

To calculate an upper limit on the branching ratio we used:
\begin{equation}
  \label{eq:ul}
{\mathrm Br(B^-\rightarrow K^-K^-\pi^+)} \leq {N^{90}\over \epsilon~ 
    N_{\mathrm B}}~~~, 
\end{equation}
where $N_{\mathrm B}$ is the number of
charged B mesons in the sample and $\epsilon$ is the efficiency for
Monte Carlo simulated events of the process B$^-$\to\Km\Km\Pip\
to survive the selection procedure. With 4.41 million hadronic \Zz\ decays,
using Br(b\to B$^\pm$) =
0.397$^{+0.018}_{-0.022}$ and $\GbbGhad = 0.2170\pm0.0009$
    \cite{LEPSLD}, we obtained 
$N_{\mathrm B}$=759\,800$^{+34\,600}_{-42\,200}$. 

The conservative approach when setting upper limits
is to use the model giving the lowest efficiency for the signal.
If one assumes resonance production, then the lowest
efficiency,  8.11\plm0.19\%,
is obtained when assuming the signal decay channel is via K$^{\star0}(892).$ 
The lowest efficiency for non-resonant decay, obtained with a
phase space particle distribution,
was found to be 11.3\plm0.2\%.

\section{Systematic Uncertainties}
\label{Sec:Syst}
Systematic uncertainties may arise from the limited accuracy with
which  $N_{\mathrm B}$
is known, from the uncertainty in the simulation used to
determine the efficiency and from the background estimation. 

\subsubsection*{Modelling of \dEdx}
\label{Sec:dedx}
To estimate the uncertainty arising from the modelling of the \dEdx\
selection criteria, we compared the efficiency of the \dEdx\ cuts in 
Monte Carlo
simulated events and in data. We took advantage of the abundance and
relative ease of reconstruction of
\Dstar\ mesons, and exploited them for testing 
the systematic uncertainties associated with the \dEdx\ selection
criteria.  
We searched for \Dstar\ mesons via their decay
into a D$^0$ and a \Pip, where the D$^0$ decays via a \Km\Pip. 
To enhance the signal to
background ratio we required the momentum of the \Dstar\ candidate to
be larger than 15 \GeVc; the \Dstar\ decay vertex to be at least 50
$\mu$m away from the interaction point; and the helicity angle,
$\theta^*$, between the kaon
momentum in the D rest frame and the D direction in the laboratory frame
to satisfy $\cos\theta^*<0.7$. 
Background estimation, after applying these selection criteria, was done as in
\cite{gcc}. 
To avoid possible uncertainty due to the difference in the momentum
spectrum of the \Dstar\ products with respect to the momentum spectrum
of tracks from the process B$^-$\to\Km\Km\Pip, we reweighted the
\dEdx\ selection efficiency as a function of the track momenta.
By applying the \dEdx\ criteria used to select the
kaon we obtained an efficiency of 56.2\% in data and 57.0\% in Monte
Carlo. The respective values for the pion selection efficiency are
75.0\% and 76.8\%. Combining all numbers, the relative uncertainty on
the signal efficiency associated
with the \dEdx\ cuts is estimated at 3.7\%.

\subsubsection*{Artificial neural network uncertainty}
\label{Sec:ANN}
\begin{figure}[]
   \begin{center}\hspace*{-.4cm}\mbox{
        \epsfig{file=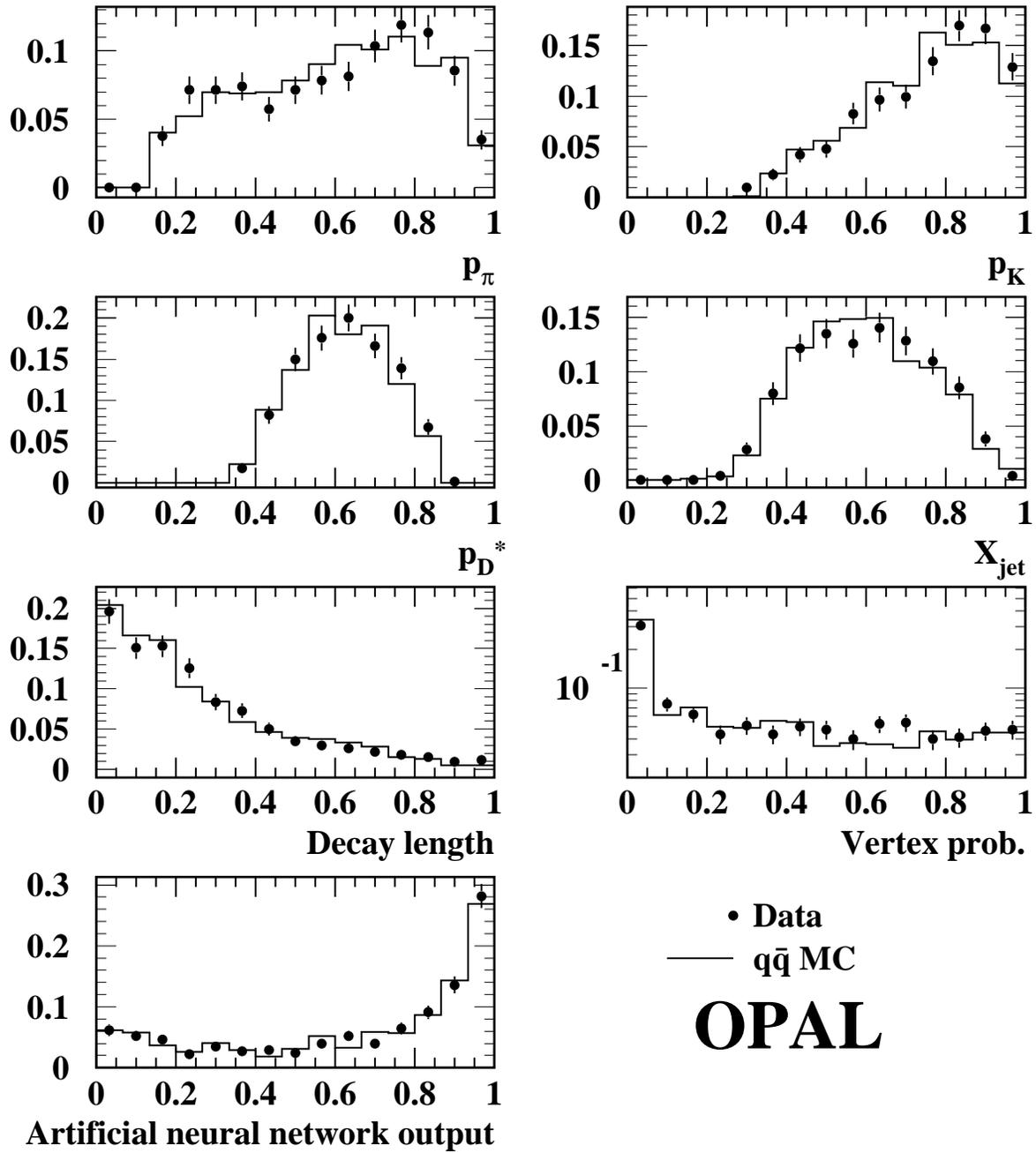}}
   \caption{Comparison of ANN input variables for \Dstar\
     candidates. The solid line represents the \qqbar\ Monte Carlo and
     the dots are the data.}
   \label{fig:Dstarnn}
\end{center}
\end{figure}

The agreement is good between the data and the \qqbar\ Monte Carlo, in all of
the input variables to the ANN (Figure
\ref{fig:nn}). However, the Monte Carlo simulation compared in that
figure represents the background and is not used in setting the upper limit.
The simulation used to set the limit is that of signal events and
thus, signal input variables should be compared. This is not possible
for the decay B$^-$\to\Km\Km\Pip. Therefore, once again we made use of
the \Dstar\ signal. We compared the kaon and
pion momentum distribution, the \Dstar\ momentum and fraction of
energy from the jet's energy, the decay length and the vertex probability.
Good agreement was achieved between the data and Monte Carlo as shown
in Figure \ref{fig:Dstarnn}. Events were reweighted as a function of
the track momenta
to reflect the signal spectrum as in the \dEdx\ uncertainty section,
and the ANN output was evaluated.
In order to assign a systematic
uncertainty to the efficiency of the ANN, we took the difference
between the fraction of \Dstar\ events passing the ANN cut in the data to that
in the \qqbar\ Monte Carlo. 
We obtained an uncertainty of 4.1\%. In addition, we took the
difference in the ANN efficiency obtained for the signal involving 
different \Kstar\ resonances. 
This difference was found to be 2.1\% and the overall
uncertainty assigned for this source was 4.6\%

\subsubsection*{B hadron lifetime and decay multiplicity}
The probability to reconstruct the signal B$^-$ meson from also
depends on the efficiency to reconstruct secondary
vertices in both hemispheres. This in turn is sensitive to the
charged decay multiplicity and lifetime of the B hadrons. The
Monte Carlo was reweighted to reflect the measured multiplicities and
lifetimes \cite{LEPSLD}. The uncertainty on these
figures gave an uncertainty of 1.1\% and 1.3\%,
respectively, on the selection efficiency.

\subsubsection*{Detector modelling}
The resolution of the tracking devices has an
effect on the efficiency.
The simulated resolutions were varied by $\pm 10\thinspace\%$
relative to the values that optimally describe the
data following the studies in~\cite{Rbmain}.
The analysis was repeated and the efficiency estimation was recalculated. 
This source contributed an uncertainty of 1.2\%.

\subsubsection*{Fragmentation modelling}
The heavy-quark fragmentation was simulated using the
function of Peterson~\etal~\cite{peterson}.
The heavy-quark fragmentation model parameter
was varied to change the mean scaled
energy of weakly-decaying bottom hadrons within the
experimental range: 
$\langle x_{E}\rangle_{\rm b}=0.702\pm 0.008$ \cite{LEPSLD}. This
change resulted in a 2\% change in the efficiency.
In addition, the heavy-quark fragmentation model was changed  
to that suggested by Collins and
Spiller~\cite{CAS} and to that of Kartvelishvili~\etal~\cite{KAV},
with parameters tuned according to Reference~\cite{AJM}.
No significant change in the resulting efficiency was observed.

\subsubsection*{Background estimation uncertainty}
The uncertainty on the fitted shape parameters and on the normalisation
gave an uncertainty on the background estimate. The different
techniques used to estimate the background resulted in consistent 
estimates with a small standard deviation (about 1 event). However,
since we used the lowest background estimate, these uncertainties were
not taken into account as they were smaller than the difference between
the mean background estimate and the one used.

\section{Results}

Combining all 
sources of systematic
uncertainties mentioned above, as well as the statistical uncertainty
in determining the efficiency and the uncertainty on  $N_{\mathrm
  B^-},$ we obtained an uncertainty of 8.4\% on the denominator of
Equation \ref{eq:ul}. This uncertainty was incorporated according to
the method outlined in reference \cite{Cousins}.
With $N^{90}_{\mathrm res.} = 7.8$ events and $N^{90}_{\mathrm
  no~res.} = 7.4$ events we obtained:
\begin{eqnarray*}
{\mathrm Br(B^-\rightarrow K^-K^-\pi^+)}& &\leq 1.29\times 10^{-4}~~@~
90\%~{\mathrm C.L.} \\
{\mathrm Br(B^-\rightarrow K^-K^-\pi^+)}&{\mathrm~non-resonance} 
&\leq 8.79\times 10^{-5}~~@~ 90\%~{\mathrm C.L.}
\end{eqnarray*}

\section{Summary}
We have searched for the decay of charged B mesons to \Km\Km\Pip.
This decay channel is strongly suppressed in the Standard Model, but
may be large in R-parity violating models. Hence, this decay mode
may serve as a probe for new physics beyond the Standard Model. 
No evidence has been observed for such a decay. Upper limits on the 
branching ratio have been set of
$1.29\times 10^{-4},$ or of $8.79\times 10^{-5}$  if one assumes that
the decay is not via a \Kstar\ resonance, both at 90\% confidence level. 

Using these limits, and the estimate ${{\mathrm B^-\rightarrow
    K^-K^-\pi^+\over b\rightarrow ss\overline{d}}}\approx {1\over4}$
\cite{PS}, we can put new limits on the contribution of R-parity
violating couplings in this process. Starting from Equation 9 of
reference \cite{PS}, 
$$\Gamma={ m_{\mathrm b}^5f^2_{\mathrm QCD}\over 512(2\pi)^3}
\left(\left\vert\Sigma_{\mathrm n=1}^3{1\over m^2_{\tilde\nu_{\mathrm n}}}
\lambda^\prime_{\mathrm n32}\lambda^{\prime\star}_{\mathrm n21}\right\vert^2 + 
\left\vert\Sigma_{\mathrm n=1}^3{1\over m^2_{\tilde\nu_{\mathrm n}}}
\lambda^\prime_{\mathrm n12}\lambda^{\prime\star}_{\mathrm
  n23}\right\vert^2\right),$$
where $m{\mathrm _b}$ is the mass of the b quark, $f_{\mathrm QCD} =
(\alpha_s(m_{\mathrm b})/\alpha_s(m_{\tilde\nu_{\mathrm n}}))^{24/23}$,
$m_{\tilde\nu_{\mathrm n}}$ is the mass of
the sneutrino involved and $\lambda^\prime$ is a dimensionless coupling.
As an example, with  $m{\mathrm _b}=$ 4.5 \GeVcc, $f_{\mathrm QCD}
\simeq 2$, 
$m_{\tilde\nu_{\mathrm n}}$ = 100 \GeVcc\ as in \cite{PS} and
$\tau_{\mathrm B^-} = 1.65~{\mathrm ps}$ we obtain:
$${\mathrm
  \sqrt{\vert\Sigma_{n=1}^3\lambda^\prime_{n32}\lambda^{\prime\star}_{n21}
\vert^2 + 
  \vert\Sigma_{n=1}^3\lambda^\prime_{n12}\lambda^{\prime\star}_{n23}\vert^2}
<5.9\times10^{-4}},$$
which can be compared to the existing limit of 0.1 obtained from 
b\to\ ss${\mathrm\overline d}$ \cite{PS} and neutrino mass calculation \cite{PSXV}.

\section{Acknowledgement}
The authors would like to acknowledge Prof. Paul Singer for valuable
discussion related to the analysis presented here.
We particularly wish to thank the SL Division for the efficient operation
of the LEP accelerator at all energies
 and for their continuing close cooperation with
our experimental group.  We thank our colleagues from CEA, DAPNIA/SPP,
CE-Saclay for their efforts over the years on the time-of-flight and trigger
systems which we continue to use.  In addition to the support staff at our own
institutions we are pleased to acknowledge the  \\
Department of Energy, USA, \\
National Science Foundation, USA, \\
Particle Physics and Astronomy Research Council, UK, \\
Natural Sciences and Engineering Research Council, Canada, \\
Israel Science Foundation, administered by the Israel
Academy of Science and Humanities, \\
Minerva Gesellschaft, \\
Benoziyo Center for High Energy Physics,\\
Japanese Ministry of Education, Science and Culture (the
Monbusho) and a grant under the Monbusho International
Science Research Program,\\
Japanese Society for the Promotion of Science (JSPS),\\
German Israeli Bi-national Science Foundation (GIF), \\
Bundesministerium f\"ur Bildung, Wissenschaft,
Forschung und Technologie, Germany, \\
National Research Council of Canada, \\
Research Corporation, USA,\\
Hungarian Foundation for Scientific Research, OTKA T-029328, 
T023793 and OTKA F-023259.\\

\end{document}